%% file: dmb_isit09.tex
\documentclass[conference]{IEEEtran-v17}
\pdfoutput=1 

\usepackage{xspace}
\usepackage[nosort]{cite}        
\usepackage{url} 
\usepackage[cmex10]{amsmath}
\usepackage{bbm}
\usepackage{graphicx}
\usepackage{paralist}
\usepackage{fancyref}
\usepackage[stretch=16,shrink=16,step=4]{microtype}
\newlength{\figwidth}
\makeatletter
\def\@IEEEinterspaceratioM{0.265}
\def\@IEEEinterspaceMINratioM{0.1651}
\def\@IEEEinterspaceMAXratioM{0.38}

\def\@IEEEinterspaceratioB{0.31}
\def\@IEEEinterspaceMINratioB{0.19}
\def\@IEEEinterspaceMAXratioB{0.38}
\@IEEEtunefonts
\makeatother

\input{vmr-symbols-vecbold.tex}
\input{standard-macros.tex}

\input{defs.tex}

\begin{document}
\IEEEoverridecommandlockouts

\title{On the Sensitivity of Noncoherent Capacity to the Channel Model}



\author{
\IEEEauthorblockN{Giuseppe Durisi, Veniamin I. Morgenshtern, and  Helmut B\"{o}lcskei }
\IEEEauthorblockA{Communication  Technology Laboratory\\
 ETH Zurich, 8092 Zurich, Switzerland\\
E-mail: \{gdurisi, vmorgens, boelcskei\}@nari.ee.ethz.ch\\} 
}

\maketitle

\begin{abstract}
The noncoherent capacity of stationary discrete-time fading channels is known to be very sensitive to the fine details of the channel model. More specifically, the measure of the set of harmonics where the power spectral density of the fading process is nonzero determines if capacity grows logarithmically in SNR or slower than logarithmically. 
An engineering-relevant problem is to characterize the SNR value at which this sensitivity starts to matter.
In this paper, we consider the general class of continuous-time Rayleigh-fading channels that satisfy the wide-sense stationary uncorrelated-scattering (WSSUS) assumption and are, in addition, underspread.  For this class of channels, we show that the noncoherent capacity is close to the AWGN capacity for all SNR values of practical interest, independently of whether the scattering function is compactly supported or not. As a byproduct of our analysis, we obtain an information-theoretic pulse-design criterion for orthogonal frequency-division multiplexing systems.  
\end{abstract}
\IEEEpeerreviewmaketitle
\section{Introduction and Summary of Results}
\label{sec:introduction}

The capacity of fading channels in the absence of channel state information (CSI) both at the transmitter and the receiver\footnote{This quantity, under the additional assumption that the transmitter and the receiver are aware of the channel law, is typically called \emph{noncoherent} capacity; in the remainder of this paper, it  will be referred to simply as capacity.} is  notoriously difficult to analyze even for simple channel models~\cite{abou-faycal01-05a}. Most of the results available in the literature pertain to either low or high signal-to-noise ratio (SNR) asymptotics. 
While in the low-SNR regime the capacity behavior seems robust with respect to the underlying channel model (see for example~\cite{durisi08-04a} for a detailed review of low-SNR capacity results), this is not the case in the high-SNR regime, where capacity is very sensitive to the \emph{fine details} of the channel model, as we are going to argue next.\looseness-1

Consider, as an example, a discrete-time stationary frequency-flat time-selective Rayleigh-fading channel subject to additive white Gaussian noise (AWGN). 
Here, the channel law is fully specified by  the power spectral density (PSD)~$\psdonedp$, $\specparam \in [-1/2,1/2)$, of the fading process and by the noise variance. 
The high-SNR capacity of this channel depends on the measure~\measure of the set of harmonics~\specparam where the PSD is nonzero. 
More specifically, let~\SNR denote the SNR; if $\measure<1$, capacity behaves as~$(1-\measure)\log \SNR$, in the high-SNR regime~\cite{lapidoth05-07a}. If~$\measure=1$ and the fading process is regular, i.e.,~$\int_{-1/2}^{1/2} \log \psdonedp d\specparam> -\infty$, then the high-SNR capacity behaves as~$\log\log \SNR$~\cite{lapidoth05-07a}. As a consequence, two channels, one with PSD equal to~$1/\spreadoned$ for~$\specparam \in [-\spreadoned/2,\spreadoned/2]$ ($0<\spreadoned<1$) and~$0$ else, and the other one with PSD equal to~$(1-\uncertainty)/\spreadoned$ for~$\specparam \in [-\spreadoned/2,\spreadoned/2]$ and~$\uncertainty/(1-\spreadoned)$ else ($0<\uncertainty<1$), will have completely different high-SNR capacity behavior, no matter how small~$\uncertainty$ is. 
A result like this is clearly unsatisfactory from an engineering viewpoint, as the measure of the support of a PSD cannot be determined through channel measurements.
Such a sensitive dependency of the capacity behavior on the fine details of the channel model (by fine details here, we mean details that, in the words of Slepian~\cite{slepian76-03a}, have ``\dots no direct meaningful counterparts in the real world \dots ''), should make one question the validity of the channel model itself.\looseness-1

An engineering-relevant problem is then to determine the SNR value at which capacity starts being sensitive to such fine details. 
An attempt to resolve this problem was recently made in~\cite{etkin06-04a}, where, for a first-order Gauss-Markov fading process, the SNR beyond which capacity behaves  as~$\log\log \SNR$ is computed as a function of the innovation rate of the process. The main limitation of this result is that it is based on a very specific channel model and that it is difficult to link the innovation rate to physical channel parameters.

In this paper, we attempt to address the problem in more generality.
Rather than focusing on a specific discretized channel model, we start from the general class of continuous-time Rayleigh-fading linear time-varying (LTV) channels that satisfy the \emph{wide-sense stationary} (WSS) and \emph{uncorrelated scattering} (US) assumptions~\cite{bello63-12a} and that are, in addition, \emph{underspread}~\cite{kennedy69}. The Rayleigh-fading and the WSSUS assumptions imply that the statistics of the channel are fully characterized by its \emph{scattering function}~\cite{bello63-12a}; the underspread assumption is satisfied if the scattering function is highly concentrated around the origin of the Doppler-delay plane. More concretely, we shall say that a WSSUS channel is underspread if  its scattering function has only a fraction~$\uncertainty\ll 1$ of its volume outside a rectangle of area~$\spread \ll 1$ (see Definition~\ref{dfn:underspread} in the next section). Our main result is the following: 
 We provide a lower bound on the capacity of continuous-time WSSUS underspread Rayleigh-fading channels that is explicit in the parameters~\spread and~\uncertainty. 
On the basis of this bound, we show that for all SNR values~\SNR that satisfy $\sqrt{\spread}\ll \SNR \ll 1/(\spread+\epsilon)$, the fading-channel capacity is close to the capacity of a nonfading AWGN channel with the same SNR. Hence, the fading-channel capacity grows logarithmically in SNR  up to (high) SNR values~$\SNR \ll 1/(\spread+\epsilon)$.

A crucial step in the derivation of our capacity lower bound is the discretization of the continuous-time channel input-output (I/O) relation, which is accomplished by transmitting and receiving on an \emph{orthonormal Weyl-Heisenberg} (WH) set~\cite[Ch. 8]{christensen03a} of time-frequency shifts of a pulse~\logonp. The resulting signaling scheme can be interpreted as pulse-shaped (PS) orthogonal frequency-division multiplexing (OFDM). 
A similar approach was used in~\cite{durisi08-04a} to characterize the capacity of WSSUS underspread fading channels in the low-SNR regime. Differently from~\cite{durisi08-04a}, in this paper, we explicitly account for the intersymbol and intercarrier interference terms in the discretized I/O relation. This is crucial, as unlike in the low-SNR regime, these terms play a fundamental role at high SNR. 
Finally, as an interesting byproduct of our analysis, we obtain an  information-theoretic pulse-design criterion for PS-OFDM systems that operate over  WSSUS underspread fading channels.

\subsubsection*{Notation}

 Uppercase boldface letters denote matrices, and lowercase boldface letters
 designate vectors. The Hilbert space of complex-valued finite-energy signals is denoted as~\hilfunspacep and~$\vecnorm{\cdot}$ stands for the norm in~\hilfunspacep. The set of positive real numbers is denoted as~\positivereals; the superscript~$\tp{}$ stands for transposition,  ~$\Ex{}{\cdot}$ is the expectation operator, and~$\four[\cdot]$ stands for the Fourier transform operator. For two vectors~\veca and~\vecb of equal dimension, the Hadamard product is denoted as~$\veca \had \vecb$, and for two functions~$f(x)$ and~$g(x)$, the notation~$f(x)=\landauo(g(x))$ means that~$\lim_{x\to 0} f(x)/g(x)=0$. Finally,~$\krond[k]$ is defined as~$\krond[0]=1$ and~$\krond[k]=0$ for all~$k\neq 0$.


\section{System Model} 
\label{sec:system_model}

\subsection{The Continuous-Time Input-Output Relation} 
\label{sec:the_continuous_time_input_output_relation}

In the following, we briefly summarize the continuous-time WSSUS underspread Rayleigh-fading channel model employed in this paper. For a more complete description of this model, the interested reader is referred to~\cite{durisi08-04a}.  
The I/O relation of a continuous-time stochastic LTV channel~$\CHop$ can be written as
\ba
	\outp(\time) &= (\CHop\inp)(\time) +\wgn(\time) \notag \\
	&=\int\!\tvirp\inp(\time-\delay)d\delay +\wgn(\time)
\label{eq:ltv-io}
\ea
where~\outpp is the received signal.
As in~\cite[Model 2]{wyner66-03a}, the stochastic transmit signal~\inpp belongs to the subset $\hilfuntimeband\subset\hilfunspacep$ of signals that are approximately limited to a duration of~$\duration \sec$ and strictly limited to a bandwidth of $\bandwidth \Hz$; furthermore,~\inpp satisfies the average-power constraint $(1/\duration)\Ex{}{\vecnorm{\inpp}^2}\leq\Pave$. The signal~\wgnp is a zero-mean unit-variance proper AWGN process, and the channel impulse response~\tvirp is a zero-mean jointly proper Gaussian (JPG) process that satisfies the WSSUS assumption
%
%
\be
	\Ex{}{\tvirp \conj{\tvir}(\time',\delay')}=\corrtd(\time-\time',\delay)\dirac(\delay-\delay').
\een
Hence, the time-delay correlation function~\corrtdp, or, equivalently, the Doppler-delay scattering function~$\scafunp\define\four_{\time \to \doppler}\{\corrtdp\}$ fully characterizes the channel statistics. In the remainder of the paper, we let the scattering function be normalized in volume according to
%
%
	$\spreadint{\scafunp}=1$.
%
As we assumed unit-variance noise, the SNR is given by~$\SNR=\Pave/\bandwidth$. 
Even though~\inpp has bandwidth no larger than~\bandwidth, the signal $(\CHop \inp)(\time)$ might not satisfy a strict bandwidth constraint. For simplicity of exposition, we assume that \outpp in~\eqref{eq:ltv-io} is passed through an ideal filter of bandwidth~\bandwidth, so that both~\inpp and~\outpp are strictly limited to a bandwidth of $\bandwidth \Hz$. 
The capacity of the resulting effective channel can be upper-bounded by $\cawgnp=\bandwidth \log(1+\SNR)$, which is the capacity of a nonfading AWGN channel with the same SNR~\cite{wyner66-03a}.

\subsubsection*{A Robust Definition of Underspread Channels} 
\label{sec:a_robust_definition_of_underspread_channels}
Qualitatively speaking, WSSUS underspread channels are WSSUS channels with a scattering function that is highly concentrated around the origin of the Doppler-delay plane~\cite{bello63-12a}.
A mathematically precise definition of the underspread property is available for the case where~\scafunp is {\it compactly supported} within a rectangle. In this case, the channel is said to be underspread if the support area of~\scafunp is much smaller than~$1$ (see for example~\cite{kozek97a,durisi08-04a}).
The compact-support assumption, albeit mathematically convenient, is a fine detail of the channel model in the terminology introduced in the previous section, because it is not possible to determine through channel measurements whether~\scafunp is indeed compactly supported or not. 
However, the results discussed in the previous section hint at a high sensitivity of capacity to this fine detail. To better understand and  quantify this sensitivity, we need to take a more general approach. 
We replace the compact-support assumption by the following more robust and physically meaningful assumption: \scafunp has a small fraction of its total volume outside a rectangle of an area that is much smaller than~$1$. More precisely, we have the following definition.
\begin{dfn}\label{dfn:underspread}
    Let~$\maxDoppler,\maxDelay \in \positivereals, \epsilon \in[0,1]$, and let~\chsetp be the set of all Rayleigh-fading WSSUS channels~$\CHop$ with scattering function~\scafunp satisfying
	\be
	\label{eq:underspread_definition}
		\int_{-\maxDoppler}^{\maxDoppler}\int_{-\maxDelay}^{\maxDelay} \scafunp d\delay d\doppler  \geq 1-\uncertainty.
	\ee
	We say that the channels in~\chsetp are \emph{underspread} if~$\spread=4\maxDelay\maxDoppler \ll 1$ and~$\uncertainty\ll 1$.
\end{dfn}

Typical wireless channels are (highly) underspread, with most of the volume of~\scafunp supported over a rectangle of area $\spread \approx 10^{-3}$ for land-mobile channels, and~\spread as small as~$10^{-7}$ for certain indoor channels with restricted terminal mobility. Note that~$\uncertainty=0$ in~\fref{dfn:underspread} yields the compact-support underspread definition of~\cite{kozek97a,durisi08-04a}. 

It is now appropriate to provide a preview of the nature of the results we are going to obtain on the basis of the novel underspread definition just introduced.  We will show that, as long as~$\spread \ll 1$ and~$\uncertainty \ll 1$, the capacity of all channels in~\chsetp, independently of whether their scattering function is compactly supported or not, is close to the AWGN capacity~\cawgn for all SNR values typically encountered in practical wireless communication systems. To establish this result, we choose a specific transmit and receive scheme (detailed in the next section), which yields a capacity lower bound that is close to the upper bound~\cawgn.

\section{A Lower Bound on Capacity} 
\label{sec:a_lower_bound_on_capacity}

\subsection{Discretization of the Input-Output Relation} 
\label{sec:discretization_of_the_input_output_relation}

The starting point for an information-theoretic analysis of the continuous-time problem under consideration is the discretization of the I/O relation~\eqref{eq:ltv-io}. This is accomplished by transmitting and receiving on the highly structured WH set~$\WHset\define\left\{\slogonp=\logon(\time-\dtime\tstep)\cex{\dfreq\fstep\time}\right\}_{\dtime,\dfreq \in \integers}$ 
of time-frequency shifts of the pulse~\logonp. We choose~\logonp, \tstep, and~\fstep such that the following properties are satisfied:
\begin{inparaenum}[i)]
	\item \logonp has unit energy, is strictly bandlimited, and satisfies $\lim_{\abs{\time}\to\infty}\abs{\time}^{1+\eta}\logonp=0$, for some $\eta>0$;
	\item the signals in the WH set \WHset are orthonormal;
	\item $\slogonp \in \hilfuntimeband$, for~$\allz{\dtime}{\tslots},\allz{\dfreq}{\fslots}$,  where $\tslots=\duration/\tstep-c$ and~$\fslots=\bandwidth/\fstep$, with~$c$ being a constant that depends on the pulse~\logonp, but not on~\duration.
\end{inparaenum}
An explicit construction of a family of WH sets~\WHset for which Properties~i)-iii) are satisfied is provided in~\fref{sec:a_simple_pulse}.
We consider  transmit signals of the form
\be
\label{eq:canonical-input}
 	\inpp=\sum_{\dtime=0}^{\tslots-1}\sum_{\dfreq=0}^{\fslots-1}
 		\inp[\dtdf] \slogon(\time)
\ee
where the data symbols~$\inp[\dtdf] \in \complexset$ are chosen such that the power constraint~$(1/\duration)\Ex{}{\vecnorm{\inpp}^2}\leq\Pave$ is satisfied, i.e.,
\be
 \label{eq:apc}
  \sum_{\dtime=0}^{\tslots-1}\sum_{\dfreq=0}^{\fslots-1}
 		\Ex{}{\abs{\inp[\dtdf]}^{2}} \le \tslots\tstep\Pave .
\ee
Properties~i)-iii) guarantee that~\inpp in~\eqref{eq:canonical-input} belongs to~\hilfuntimeband as detailed in~\cite{durisi09-01b}.
The received signal~\outpp is projected onto the signal set $\left\{g_{k,n}(t)\right\}_{k=0,n=0}^{K-1,N-1}$ to obtain 
\ba
	&\underbrace{\inner{\outp}{\slogon}}_{\define\,\outp[\dtdf]} = 
	\underbrace{\inner{\CHop\slogon}{\slogon}}_{\define\,\ch[\dtdf]} \inp[\dtdf]\notag\\
	&\quad\qquad\qquad+\mathop{\sum_{\altdtime=0}^{\tslots-1}\sum_{\altdfreq=0}^{\fslots-1}}_{(\altdtdf)\neq(\dtdf)} \underbrace{\inner{\CHop\logon_{\altdtdf}}{\slogon}}_{\define\,\interfp}\inp[\altdtdf] +\underbrace{\inner{\wgn}{\slogon}}_{\define\,\wgn[\dtdf]}\notag\\
	&=\ch[\dtdf]\inp[\dtdf]+\mathop{\sum_{\altdtime=0}^{\tslots-1}\sum_{\altdfreq=0}^{\fslots-1}}_{(\altdtdf)\neq(\dtdf)} \interfp\inp[\altdtdf]
	+\wgn[\dtdf]
	\label{eq:discretized I/O with interference}
\ea
for each \emph{time-frequency} slot~$(\dtime,\dfreq)$, $\allz{\dtime}{\tslots},\, \allz{\dfreq}{\fslots}$. We refer to the channel with I/O relation~\eqref{eq:discretized I/O with interference} as the discretized channel \emph{induced} by the  WH set~\WHset. 
Note that we do not require that all signals in~\hilfuntimeband satisfying $(1/\duration)\Ex{}{\vecnorm{\inpp}^2}\leq\Pave$ can be represented in the form~\eqref{eq:canonical-input}. As a consequence, the capacity of the channel~\eqref{eq:discretized I/O with interference}, defined in the next section, is a lower bound on the capacity of the underlying continuous-time channel~\eqref{eq:ltv-io}.
The second term in~\eqref{eq:discretized I/O with interference} corresponds to intersymbol and intercarrier interference. The variance of this term depends on the time-frequency localization properties of~\logonp~\cite{durisi09-01b}. 
The orthonormality of~\WHset implies that~$\wgn[\dtdf]$ in~\eqref{eq:discretized I/O with interference} is \iid $\jpg(0,1)$. 
A necessary condition for orthonormality of the set~\WHset is $\tstep \fstep\ge1$ \cite[Cor. 7.5.1 and Cor. 7.3.2]{groechenig01a}, and good time-frequency localization of the signals in the orthonormal set~\WHset is possible for~$\tfstep>1$~\cite[Th. 4.1.1]{christensen03a}. 
Finally, we note that discretizing the continuous-time I/O relation by transmitting and receiving on WH sets ensures that the induced discretized channel~\eqref{eq:discretized I/O with interference} inherits the (two-dimensional) stationarity properties of the underlying continuous-time channel~\cite{durisi09-01b}, a fact that is crucial for the ensuing analysis.

\subsection{I/O Relation in Vector-Matrix Form} 
\label{sec:i_o_relation_in_vector_matrix_form}

For each~$\dtime \in \{0,1,\ldots,\tslots-1\}$, we arrange the data  symbols~$\inp[\dtdf]$, the received signal samples~$\outp[\dtdf]$, the channel coefficients~$\ch[\dtdf]$, and the noise samples~$\wgn[\dtdf]$ in corresponding vectors. For example, the \fslots-dimensional vector that contains the data symbols in the $\dtime$th time slot is defined as
\be
	\inpvec[\dtime]\define\tp{\mat\inp[\dtime,0]\; \inp[\dtime,1]\; \cdots\;
		\inp[\dtime,\fslots-1]\emat}.
\een
To obtain a compact notation, we further stack~\tslots contiguous $\fslots$-dimensional input, output, channel, and noise vectors, into  corresponding $\tslots \fslots$-dimensional vectors~\inpvec, \outpvec, \chvec, and~\wgnvec, respectively.  For the data symbols, for example, this
results in the $\tslots\fslots$-dimensional vector
%
$	\inpvec\define\tp{\mat\tp{\inpvec}[0]\; \tp{\inpvec}[1]\; \cdots\;
		\tp{\inpvec}[\tslots-1]\emat}$.
%
Finally, we arrange the intersymbol and intercarrier interference terms~$\{\interfp\}$  in a~$\tslots\fslots \times \tslots\fslots$ matrix~\interfmat with entries
%
$[\interfmat]_{\altdfreq+\altdtime\fslots,\dfreq+\dtime\fslots}=\interfp$ if~$(\altdtdf) \neq (\dtdf)$ and $0$ else.
With these definitions, we can now compactly express the I/O relation~\eqref{eq:discretized I/O with interference} as
\be
\label{eq:I_O_in_vector_matrix_form}
	\outpvec=\chvec\had\inpvec+ \interfmat\inpvec +\wgnvec.
\ee
%
%


\subsection{Definition of Capacity} 
\label{sec:capacity_definition}

For a given  WH set~\WHset satisfying Properties i)-iii) in \fref{sec:discretization_of_the_input_output_relation} and a given continuous-time channel $\CHop$, the capacity of the induced discretized channel~\eqref{eq:discretized I/O with interference} is given by~\cite{durisi09-01b}
\bas
 	\capacityp\define\lim_{\tslots\to\infty} \frac{1}{\tslots\tstep} \sup_{\setinp}
		\mi(\outpvec;\inpvec).
\eas
Here, the supremum is taken over the set~\setinp of all distributions on~\inpvec
that satisfy  the average-power
constraint~\eqref{eq:apc}.
As already mentioned, \capacity is a lower bound on the capacity of the continuous-time channel~\eqref{eq:ltv-io}.


\subsection{The Capacity Lower Bound} 
\label{sec:the_lower_bound}

\begin{thm}\label{thm:lower bound}
	
Let~\WHset  be a WH set satisfying Properties i)-iii) in \fref{sec:discretization_of_the_input_output_relation} and consider an arbitrary Rayleigh-fading WSSUS channel in the set~\chsetp. Then, for a given SNR~\SNR and  a given bandwidth~\bandwidth, and under the technical condition\footnote{This technical condition is not restrictive for underspread channels if \tstep and \fstep are chosen so that~$\maxDoppler\tstep=\maxDelay\fstep$ (see \fref{sec:reduction_to_a_square_setting}). In this case,~$2\maxDoppler\tstep =\sqrt{\spread\tfstep}\ll 1$  for all values of~\tfstep of practical interest.}~$\altspread\define2\maxDoppler\tstep< 1$, the capacity of the discretized channel induced by~\WHset is lower-bounded as:
	\be
	\label{eq:capacity_lower_bound_explicit_in_spread_and_uncertainty}
	\bs
	\capacity(\SNR) \ge&\,\LBsimplep\\
	=&\, \frac{\bandwidth}{\tfstep}\Biggl\{ \Ex{\ch}{	\log\lefto( 1 + \frac{\tfstep\SNR(1-\uncertainty) \minambsquare\! \abs{\ch}^{2}}{1 + \tfstep\SNR(\maxsumambsquare+\uncertainty) } \right)} \\
	&- \inf_{0<\noisesplit<1}\Biggl[ \altspread \log\lefto(1 + \frac{\tfstep\SNR}{\noisesplit\altspread}\right)\\
	&+
	(1-\altspread)\log\lefto(1 +\frac{\tfstep\SNR\,\epsilon}{\noisesplit(1-\altspread)}\right)\\
	&+\log\lefto(1 +\frac{\tfstep\SNR}{1-\noisesplit}(\maxsumambsquare +\uncertainty) \right)\Biggr]\Biggr\}
	\es
	\ee
	where~$\ch\distas\jpg(0,1)$, $\displaystyle	\minambsquare\define\min_{(\doppler,\delay) \in \spreadset} \abs{\afp}^2$, 
	\be
	\maxsumambsquare\define\max_{(\doppler,\delay) \in \spreadset} \mathop{\sumdtime\sumdfreq}_{(\dtdf)\neq(0,0)}\abs{\af_{\logon}(\doppler -\dfreq\fstep,\delay -\dtime\tstep)  }^{2},
	\een
	with $\spreadset \define [-\maxDoppler,\maxDoppler] \times [-\maxDelay,\maxDelay]$, and where~$\afp\define \int\logon(\time)\conj{\logon}(\time-\delay)
	 		\cexn{\doppler\time} d\time$ denotes the ambiguity function of~\logonp.
\end{thm}

\paragraph{A Glimpse of the Proof}

As the proof of~\fref{thm:lower bound} is rather involved, we only provide a summary of the main steps, leaving the details to~\cite{durisi09-01b}. We first obtain a lower bound on~\capacity by assuming a specific distribution on~\inpvec that satisfies~\eqref{eq:apc}, namely by taking~$\inp[\dtdf]$ to be \iid $\jpg(0,\tfstep\SNR)$. The chain rule for mutual information and the nonnegativity of mutual information yield\looseness-1
\be
	\mi(\outpvec;\inpvec) 
	\ge \mi(\outpvec;\inpvec\given\chvec) -
		\mi(\outpvec;\chvec\given\inpvec).
\een
The first term in the lower bound~\eqref{eq:capacity_lower_bound_explicit_in_spread_and_uncertainty} is obtained from~$\mi(\outpvec;\inpvec\given\chvec)$, by treating the interference term~$\interfmat\inpvec$ in~\eqref{eq:I_O_in_vector_matrix_form} as additional noise and using the fact that Gaussian noise is the worst noise when~\inpvec is JPG and~\chvec is known at the receiver~\cite[Lemma II.2]{diggavi01-11a}. The other terms in~\eqref{eq:capacity_lower_bound_explicit_in_spread_and_uncertainty} are a result of upper-bounding~$\mi(\outpvec;\chvec\given\inpvec)$ as follows. Let~$\wgnvecone\distas\jpg(\veczero,\noisesplit\matI)$ and~$\wgnvectwo\distas\jpg(\veczero,(1-\noisesplit)\matI)$, where $0<\noisesplit<1$, be~$\tslots\fslots$-dimensional independent JPG vectors. Furthermore, let~$\outpvecone=\chvec\had \inpvec +\wgnvecone$ and~$\outpvectwo=\interfmat\inpvec +\wgnvectwo$. 
By the data-processing inequality and the chain rule for mutual information, we have that
\be
	\mi(\outpvec;\chvec\given \inpvec) \leq \mi(\outpvecone,\outpvectwo; \chvec \given \inpvec) = \mi(\outpvecone;\chvec\given \inpvec) 
	+ \mi(\outpvectwo;\chvec \given \inpvec,\outpvecone).
\een
The second and the third term in the lower bound~\eqref{eq:capacity_lower_bound_explicit_in_spread_and_uncertainty} now follow from~$\mi(\outpvecone;\chvec\given \inpvec)$ by direct application
of~\cite[Th.~3.4]{miranda00-02a}, which is an extension of Szeg\"o's theorem 
(on the asymptotic eigenvalue distribution of Toeplitz matrices) to two-level Toeplitz matrices, and by invoking~\eqref{eq:underspread_definition}. 
The fourth term follows from~$\mi(\outpvectwo;\chvec \given \inpvec,\outpvecone)$ through simple bounding steps involving the Hadamard and Jensen inequalities and by again invoking~\eqref{eq:underspread_definition}.
The dependency of the first and the fourth term in the lower bound on the ambiguity function~\afp is through the properties of the second-order statistics of the channel coefficients~$\ch[\dtdf]$ and~$\interfp$.
\paragraph{Remarks}
The lower bound~\LBsimple in~\eqref{eq:capacity_lower_bound_explicit_in_spread_and_uncertainty}  is not useful in the asymptotic regimes~$\SNR\to 0$ and~$\SNR \to \infty$. In fact, the bound even turns negative when~\SNR is sufficiently small or sufficiently large. Nevertheless, as shown in~\fref{sec:numerical_results}, for underspread channels,~\LBsimple evaluated for particular WH sets is close to the capacity upper bound~\cawgn over all SNR values of practical interest. 
In the next two sections, we list some  properties of~\LBsimple   (proven in~\cite{durisi09-01b}), which will be used in \fref{sec:numerical_results}.

\subsection{Reduction to a Square Setting} 
\label{sec:reduction_to_a_square_setting}

	The lower bound~\LBsimplep depends on seven parameters and is therefore difficult to analyze. We show next that if~\tstep and~\fstep are chosen so that~$\maxDoppler\tstep=\maxDelay\fstep$, a condition often referred to as \emph{grid matching rule}~\cite[Eq. (2.75)]{kozek97a}, two of these seven parameters can be dropped without loss of generality. 
\begin{lem}\label{lem:square_setting}
	Let~\WHset be a WH set satisfying Properties i)-iii) in \fref{sec:discretization_of_the_input_output_relation}. Then, for any~$\beta>0$, \pagebreak
	\begin{multline*}
	\LBsimple(\SNR,\logonp,\tstep,\fstep,\maxDelay,\maxDoppler,\uncertainty)\\=\LBsimple\lefto(\SNR,\sqrt{\beta}\logon(\beta \time),\frac{\tstep}{\beta},\beta\fstep,\frac{\maxDelay}{\beta},\beta\maxDoppler,\uncertainty\right).
	\end{multline*}
	In particular, assume that~$\maxDoppler\tstep=\maxDelay\fstep$ and let~$\beta=\sqrt{\tstep/\fstep}=\sqrt{\maxDelay/\maxDoppler}$ and~$\altlogonp=\sqrt{\beta}\logon(\beta \time)$. Then,
	\begin{multline}
	\label{eq:LB_trasformation}	\LBsimplep\\=\LBsimple\lefto(\SNR,\altlogon,\sqrt{\tfstep},\sqrt{\tfstep},\sqrt{\spread}/2,\sqrt{\spread}/2,\uncertainty\right)\\
\define\LBsquarep.
	\end{multline}
\end{lem}

In the remainder of the paper, for the sake of simplicity of exposition, we will choose~\tstep and~\fstep so that the grid matching rule~$\maxDoppler\tstep=\maxDelay\fstep$ is satisfied. Then, as a consequence of \fref{lem:square_setting}, we can (and will) only consider, without loss of generality,  WH sets of the form~\WHsetsquare and WSSUS channels in the class~$\chset(\sqrt{\spread}/2,\sqrt{\spread}/2,\uncertainty)$. 

\subsection{Pulse-Design Criterion and Approximation for~\minambsquare and~\maxsumambsquare} 
\label{sec:taylor_based_approximation_for_}

The lower bound in~\eqref{eq:capacity_lower_bound_explicit_in_spread_and_uncertainty} can be tightened by maximizing it over all  WH sets satisfying Properties i)-iii) in \fref{sec:discretization_of_the_input_output_relation}. This maximization implicitly provides an information-theoretic design criterion for~\logonp, \tstep, and~\fstep. 
Classic design rules for~\logonp available in the OFDM literature (see, for example,~\cite{matz07-05a} and references therein) are based on a maximization of the signal-to-interference ratio in~\eqref{eq:discretized I/O with interference}, for a fixed value of~\tfstep  (typically,~$\tfstep\approx 1.2$). 
The maximization of the lower bound~\eqref{eq:capacity_lower_bound_explicit_in_spread_and_uncertainty} yields a more complete picture as it explicitly reveals the interplay between the product~\tfstep and the time-frequency localization properties of~\logonp, reflected through the quantities~\minambsquare and~\maxsumambsquare. 
Unfortunately, the maximization of~\LBsimple  over~\WHset seems complicated, as the dependency of~\minambsquare and \maxsumambsquare on~\WHset is difficult to characterize analytically. This problem can be partially overcome when~$\spread \ll 1$. In this case, a first-order Taylor-series expansion of~\minambsquare and~\maxsumambsquare around~$\spread=0$ yields an accurate picture.

\begin{lem}\label{lem:derivative_based_approx}
	Let~\WHsetsquare be a WH set satisfying Properties i)-iii) in \fref{sec:discretization_of_the_input_output_relation}. Assume that~$\logonp$ is real-valued and even, and that~\afp is differentiable in the points $(\dfreq\srtfstep,\dtime\srtfstep)$ for all~$(\dfreq,\dtime)$ and twice differentiable in~$(0,0)$; let $\logonfp=\four[ \logonp] $ and define~$\spreadsquareset=[-\sqrt{\spread}/2,\sqrt{\spread}/2] \times [-\sqrt{\spread}/2,\sqrt{\spread}/2]$. For~$\spread \ll 1$, we have
	\be
	\label{eq:min_approx}
	\minambsquare=\min_{(\doppler,\delay) \in \spreadsquareset} \abs{\afp}^2= 1-\constm \spread +\landauo(\spread)
	\ee
	where~$\constm=\pi^2(\eftime^2 +\efband^2)$ with
	\be
		\eftime^2=\int\time^2 \abs{\logonp}^2 d\time, \quad
		\efband^2=\int\freq^2 \abs{\logonfp}^2 d\freq.	
	\een
	Moreover, still under the assumption that~$\spread \ll 1$, we have
	\be
	\label{eq:max_approx}
	\bs
	\maxsumambsquare&=\max_{(\doppler,\delay) \in \spreadsquareset} \mathop{\sumdtime\sumdfreq}_{(\dtdf)\neq(0,0)}\abs{\af_{\logon}(\doppler -\dfreq\fstep,\delay -\dtime\tstep)  }^{2}\\
	&=  \constM \spread+\landauo(\spread)
	\es
	\ee
	where $\displaystyle\constM=\mathop{\sumdtime\sumdfreq}_{(\dtdf)\neq(0,0)}\left[\abs{\dernup}^2+\abs{\dertaup}^2\right]/4$, with \pagebreak
	\be
	\bs
		\dernup&=-j2\pi \int \time \logonp \logon(\time+\dtime\srtfstep)\cex{\dfreq\srtfstep\time}d\time\\
		\dertaup&=j2\pi \int \freq \logonf(\freq-\dfreq\srtfstep)\logonfp\cexn{\dtime\srtfstep\freq} d\freq.
	\es
	\een
\end{lem}
%
%

\subsection{A Simple WH Set} 
\label{sec:a_simple_pulse}
We next present an example of a family of WH  sets~$\WHsetsquare$ satisfying  Properties i)-iii) in \fref{sec:discretization_of_the_input_output_relation}, and for which, in addition,~\logonp is real-valued and even.
Take~$1<\tfstep<2$, let~$\dur=\srtfstep$,~$\rolloff=\tfstep-1$, and~$\logonfp=\four\{\logonp\}$. We choose~\logonfp as the square root of a raised cosine:\looseness-1
\be
\logonf(\freq)=
\begin{cases}
\sqrt{\dur}, 
& \text{if}\quad\abs{\freq}\leq \frac{1-\rolloff}{2\dur} \\
\sqrt{\frac{\dur}{2}  (1 +\auxfun(\freq))},
& \text{if}\quad\frac{1-\rolloff}{2\dur} \leq \abs{\freq}\leq 
\frac{\dur}{2} \\
0, & \text{otherwise}
\end{cases}
\een
where~$\auxfun(\freq)=\cos\lefto[ \frac{\pi \dur}{\rolloff}\left(\abs{\freq}-\frac{1-\rolloff}{2\dur}\right)\right]$.
The signal~\logonfp has unit energy, is real-valued and even, and  satisfies 
\be
\label{eq:prop_raised_cos}
	\sumdfreq \logonf(\freq-\dfreq/\dur)\logonf(\freq-\dfreq/\dur-\dtime\dur)=\dur\krond[\dtime].
\ee
As a consequence of~\eqref{eq:prop_raised_cos}, by~\cite[Th. 8.7.2]{christensen03a}, the WH set~$(\logon,1/\srtfstep,1/\srtfstep)$  is a tight WH frame for~\hilfunspacep. Consequently,  the WH set~$(\logon,\srtfstep,\srtfstep)$ is orthonormal by~\cite[Th. 7.3.2]{groechenig01a}. Finally, we show in~\cite{durisi09-01b} that $\lim_{\abs{\time}\to \infty}\time^2\logonp=0$.

\section{Finite-SNR Analysis of the Lower Bound} 
\label{sec:numerical_results}

\begin{figure}[t]
	\centering
		\includegraphics[width=\figwidth]{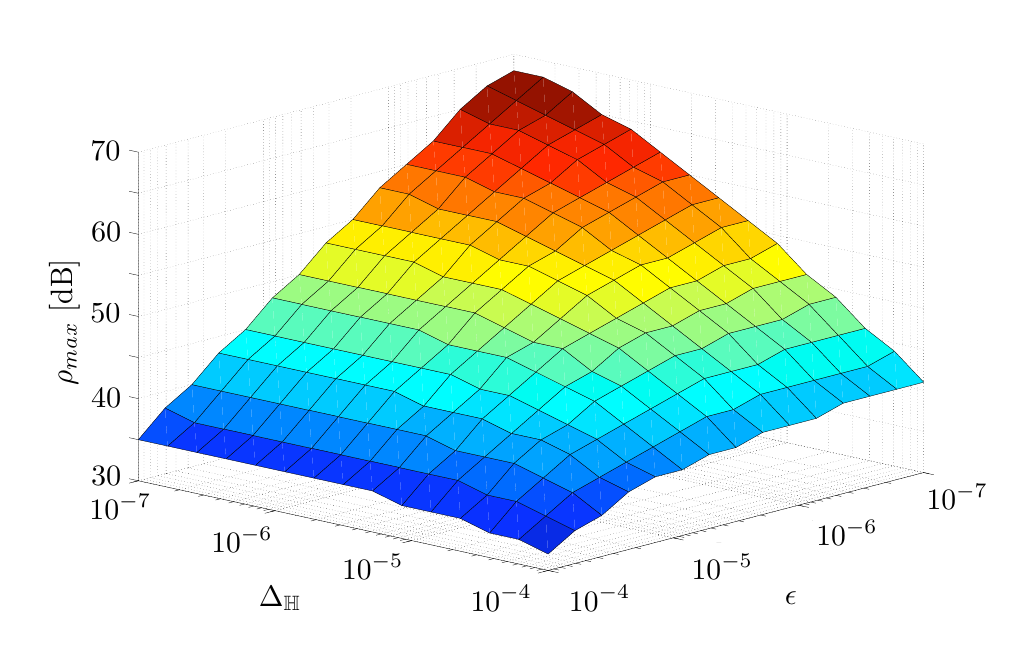}
	\caption{Maximum SNR value for which~\eqref{eq:accuracy_inequality} holds, as a function of~\spread and~\uncertainty.}
	\label{fig:max}
\end{figure}

We evaluate  the lower bound~\LBsquare  in~\eqref{eq:LB_trasformation} for the WH set constructed in the previous section, under the assumption that the underlying WSSUS channel is underspread according to~\fref{dfn:underspread}, i.e., $\spread \ll1$ and~$\uncertainty \ll 1$. More precisely, we assume~$\spread \leq 10^{-4}$ and $\uncertainty \leq 10^{-4}$.
As~$\spread \ll 1$, we can replace~\minambsquare and~\maxsumambsquare in~\LBsquare by the first-order term of their Taylor-series expansions [see~\eqref{eq:min_approx} and~\eqref{eq:max_approx}]. 
We take~$\tfstep=1.02$, which results in~$\constm\approx 25.87$ and~$\constM\approx 0.77$. 
To show that the corresponding capacity lower bound is close to the  upper bound~\cawgn for all~SNR values of practical interest, we characterize the  SNR interval~$[\SNRmin,\SNRmax]$ over which~\LBsquare is at least~$75\%$ of the AWGN capacity, i.e., 
\be
\label{eq:accuracy_inequality}
	\LBsquarepnew \geq 0.75\,\cawgnp.
\ee
The interval end points~\SNRmin and~\SNRmax can  easily be computed numerically: the corresponding values for~\SNRmax are illustrated in 
\fref{fig:max} for different~$(\spread,\uncertainty)$ pairs.
%
%
%
%
For the WH set and WSSUS underspread channels considered in this section, we have that~$\SNRmin \in [-25\dB, -7\dB]$ and~$\SNRmax \in [32\dB, 68\dB]$.

An analytic characterization of~\SNRmin and~\SNRmax is  more difficult. Insights on how these two quantities are related to the channel parameters~\spread and~\uncertainty can be obtained by replacing both sides of the inequality~\eqref{eq:accuracy_inequality} by  corresponding low-SNR approximations (to get~\SNRmin) and high-SNR approximations (to get~\SNRmax).
Under the assumption that~$\spread \leq 10^{-4}$ and~$\uncertainty \leq 10^{-4}$, this analysis, detailed in~\cite{durisi09-01b}, yields $\SNRmin\approx 13\sqrt{\spread}$ and~$\SNRmax\approx 0.22/(\spread +\uncertainty)$  for the WH set considered in this section. The following rule of thumb then holds: the capacity of all WSSUS underspread channels 
with scattering function~\scafunp having no more than~\uncertainty of its volume outside a rectangle (in the Doppler-delay plane) of area~\spread,
is close to~\cawgnp for all \SNR that satisfy $\sqrt{\spread}\ll \SNR \ll 1/(\spread + \uncertainty)$, independently of whether~\scafunp is compactly supported or not, and of its shape.
The condition $\sqrt{\spread}\ll \SNR \ll 1/(\spread + \uncertainty)$ holds for all channels and SNR values of practical interest.
%
%

%
 To conclude, an interesting open problem, the solution of which would strengthen our results, is to obtain an upper bound on the capacity of~\eqref{eq:ltv-io} based on perfect CSI at the receiver.\looseness-1
\vspace*{-1mm}
\bibliographystyle{IEEEtran}
\bibliography{IEEEabrv,publishers,confs-jrnls,ulibib,giubib}

\end{document}

%% file: vmr-symbols-vecbold.tex
\usepackage{amssymb}
\usepackage{amsfonts}
\usepackage{mathrsfs}
\usepackage{xspace}
\usepackage{bm}
\usepackage{upgreek}

\newcommand{\safemath}[2]{\newcommand{#1}{\ensuremath{#2}\xspace}}

\safemath{\bma}{\mathbf{a}}
\safemath{\bmb}{\mathbf{b}}
\safemath{\bmc}{\mathbf{c}}
\safemath{\bmd}{\mathbf{d}}
\safemath{\bme}{\mathbf{e}}
\safemath{\bmf}{\mathbf{f}}
\safemath{\bmg}{\mathbf{g}}
\safemath{\bmh}{\mathbf{h}}
\safemath{\bmi}{\mathbf{i}}
\safemath{\bmj}{\mathbf{j}}
\safemath{\bmk}{\mathbf{k}}
\safemath{\bml}{\mathbf{l}}
\safemath{\bmm}{\mathbf{m}}
\safemath{\bmn}{\mathbf{n}}
\safemath{\bmo}{\mathbf{o}}
\safemath{\bmp}{\mathbf{p}}
\safemath{\bmq}{\mathbf{q}}
\safemath{\bmr}{\mathbf{r}}
\safemath{\bms}{\mathbf{s}}
\safemath{\bmt}{\mathbf{t}}
\safemath{\bmu}{\mathbf{u}}
\safemath{\bmv}{\mathbf{v}}
\safemath{\bmw}{\mathbf{w}}
\safemath{\bmx}{\mathbf{x}}
\safemath{\bmy}{\mathbf{y}}
\safemath{\bmz}{\mathbf{z}}
\safemath{\bmzero}{\mathbf{0}}
\safemath{\bmone}{\mathbf{1}}

\bmdefine{\biad}{a}
\bmdefine{\bibd}{b}
\bmdefine{\bicd}{c}
\bmdefine{\bidd}{d}
\bmdefine{\bied}{e}
\bmdefine{\bifd}{f}
\bmdefine{\bigd}{g}
\bmdefine{\bihd}{h}
\bmdefine{\biid}{i}
\bmdefine{\bijd}{j}
\bmdefine{\bikd}{k}
\bmdefine{\bild}{l}
\bmdefine{\bimd}{m}
\bmdefine{\bind}{n}
\bmdefine{\biod}{o}
\bmdefine{\bipd}{p}
\bmdefine{\biqd}{q}
\bmdefine{\bird}{r}
\bmdefine{\bisd}{s}
\bmdefine{\bitd}{t}
\bmdefine{\biud}{u}
\bmdefine{\bivd}{v}
\bmdefine{\biwd}{w}
\bmdefine{\bixd}{x}
\bmdefine{\biyd}{y}
\bmdefine{\bizd}{z}

\bmdefine{\bixid}{\xi}
\bmdefine{\bilambdad}{\lambda}
\bmdefine{\bimud}{\mu}
\bmdefine{\bithetad}{\theta}
\bmdefine{\biphid}{\phi}

\safemath{\bmia}{\biad}
\safemath{\bmib}{\bibd}
\safemath{\bmic}{\bicd}
\safemath{\bmid}{\bidd}
\safemath{\bmie}{\bied}
\safemath{\bmif}{\bifd}
\safemath{\bmig}{\bigd}
\safemath{\bmih}{\bihd}
\safemath{\bmii}{\biid}
\safemath{\bmij}{\bijd}
\safemath{\bmik}{\bikd}
\safemath{\bmil}{\bild}
\safemath{\bmim}{\bimd}
\safemath{\bmin}{\bind}
\safemath{\bmio}{\biod}
\safemath{\bmip}{\bipd}
\safemath{\bmiq}{\biqd}
\safemath{\bmir}{\bird}
\safemath{\bmis}{\bisd}
\safemath{\bmit}{\bitd}
\safemath{\bmiu}{\biud}
\safemath{\bmiv}{\bivd}
\safemath{\bmiw}{\biwd}
\safemath{\bmix}{\bixd}
\safemath{\bmiy}{\biyd}
\safemath{\bmiz}{\bizd}

\safemath{\bmxi}{\bixid}
\safemath{\bmlambda}{\bilambdad}
\safemath{\bmmu}{\bimud}
\safemath{\bmtheta}{\bithetad}
\safemath{\bmphi}{\biphid}

\safemath{\bA}{\mathbf{A}}
\safemath{\bB}{\mathbf{B}}
\safemath{\bC}{\mathbf{C}}
\safemath{\bD}{\mathbf{D}}
\safemath{\bE}{\mathbf{E}}
\safemath{\bF}{\mathbf{F}}
\safemath{\bG}{\mathbf{G}}
\safemath{\bH}{\mathbf{H}}
\safemath{\bI}{\mathbf{I}}
\safemath{\bJ}{\mathbf{J}}
\safemath{\bK}{\mathbf{K}}
\safemath{\bL}{\mathbf{L}}
\safemath{\bM}{\mathbf{M}}
\safemath{\bN}{\mathbf{N}}
\safemath{\bO}{\mathbf{O}}
\safemath{\bP}{\mathbf{P}}
\safemath{\bQ}{\mathbf{Q}}
\safemath{\bR}{\mathbf{R}}
\safemath{\bS}{\mathbf{S}}
\safemath{\bT}{\mathbf{T}}
\safemath{\bU}{\mathbf{U}}
\safemath{\bV}{\mathbf{V}}
\safemath{\bW}{\mathbf{W}}
\safemath{\bX}{\mathbf{X}}
\safemath{\bY}{\mathbf{Y}}
\safemath{\bZ}{\mathbf{Z}}

\safemath{\bZero}{\mathbf{0}}
\safemath{\bOne}{\mathbf{1}}
\safemath{\bDelta}{\mathbf{\Delta}}
\safemath{\bLambda}{\mathbf{\UpLambda}}
\safemath{\bPhi}{\mathbf{\Upphi}}
\safemath{\bSigma}{\mathbf{\Upsigma}}
\safemath{\bOmega}{\mathbf{\Upomega}}
\safemath{\bTheta}{\mathbf{\Uptheta}}

\bmdefine{\biAd}{A}
\bmdefine{\biBd}{B}
\bmdefine{\biCd}{C}
\bmdefine{\biDd}{D}
\bmdefine{\biEd}{E}
\bmdefine{\biFd}{F}
\bmdefine{\biGd}{G}
\bmdefine{\biHd}{H}
\bmdefine{\biId}{I}
\bmdefine{\biJd}{J}
\bmdefine{\biKd}{K}
\bmdefine{\biLd}{L}
\bmdefine{\biMd}{M}
\bmdefine{\biOd}{N}
\bmdefine{\biPd}{O}
\bmdefine{\biQd}{P}
\bmdefine{\biRd}{R}
\bmdefine{\biSd}{S}
\bmdefine{\biTd}{T}
\bmdefine{\biUd}{U}
\bmdefine{\biVd}{V}
\bmdefine{\biWd}{W}
\bmdefine{\biXd}{X}
\bmdefine{\biYd}{Y}
\bmdefine{\biZd}{Z}

\bmdefine{\biDelta}{\Delta}
\bmdefine{\biLambda}{\Lambda}
\bmdefine{\biPhi}{\Phi}
\bmdefine{\biSigma}{\Sigma}
\bmdefine{\biOmega}{\Omega}
\bmdefine{\biTheta}{\Theta}

\safemath{\bimA}{\biAd}
\safemath{\bimB}{\biBd}
\safemath{\bimC}{\biCd}
\safemath{\bimD}{\biDd}
\safemath{\bimE}{\biEd}
\safemath{\bimF}{\biFd}
\safemath{\bimG}{\biGd}
\safemath{\bimH}{\biHd}
\safemath{\bimI}{\biId}
\safemath{\bimJ}{\biJd}
\safemath{\bimK}{\biKd}
\safemath{\bimL}{\biLd}
\safemath{\bimM}{\biMd}
\safemath{\bimN}{\biNd}
\safemath{\bimO}{\biOd}
\safemath{\bimP}{\biPd}
\safemath{\bimQ}{\biQd}
\safemath{\bimR}{\biRd}
\safemath{\bimS}{\biSd}
\safemath{\bimT}{\biTd}
\safemath{\bimU}{\biUd}
\safemath{\bimV}{\biVd}
\safemath{\bimW}{\biWd}
\safemath{\bimX}{\biXd}
\safemath{\bimY}{\biYd}
\safemath{\bimZ}{\biZd}

\safemath{\bimDelta}{\biDelta}
\safemath{\bimLambda}{\biLambda}
\safemath{\bimPhi}{\biPhi}
\safemath{\bimSigma}{\biSigma}
\safemath{\bimOmega}{\biOmega}
\safemath{\bimTheta}{\biTheta}

\safemath{\setA}{\mathcal{A}}
\safemath{\setB}{\mathcal{B}}
\safemath{\setC}{\mathcal{C}}
\safemath{\setD}{\mathcal{D}}
\safemath{\setE}{\mathcal{E}}
\safemath{\setF}{\mathcal{F}}
\safemath{\setG}{\mathcal{G}}
\safemath{\setH}{\mathcal{H}}
\safemath{\setI}{\mathcal{I}}
\safemath{\setJ}{\mathcal{J}}
\safemath{\setK}{\mathcal{K}}
\safemath{\setL}{\mathcal{L}}
\safemath{\setM}{\mathcal{M}}
\safemath{\setN}{\mathcal{N}}
\safemath{\setO}{\mathcal{O}}
\safemath{\setP}{\mathcal{P}}
\safemath{\setQ}{\mathcal{Q}}
\safemath{\setR}{\mathcal{R}}
\safemath{\setS}{\mathcal{S}}
\safemath{\setT}{\mathcal{T}}
\safemath{\setU}{\mathcal{U}}
\safemath{\setV}{\mathcal{V}}
\safemath{\setW}{\mathcal{W}}
\safemath{\setX}{\mathcal{X}}
\safemath{\setY}{\mathcal{Y}}
\safemath{\setZ}{\mathcal{Z}}
\safemath{\emptySet}{\varnothing}

\safemath{\colA}{\mathscr{A}}
\safemath{\colB}{\mathscr{B}}
\safemath{\colC}{\mathscr{C}}
\safemath{\colD}{\mathscr{D}}
\safemath{\colE}{\mathscr{E}}
\safemath{\colF}{\mathscr{F}}
\safemath{\colG}{\mathscr{G}}
\safemath{\colH}{\mathscr{H}}
\safemath{\colI}{\mathscr{I}}
\safemath{\colJ}{\mathscr{J}}
\safemath{\colK}{\mathscr{K}}
\safemath{\colL}{\mathscr{L}}
\safemath{\colM}{\mathscr{M}}
\safemath{\colN}{\mathscr{N}}
\safemath{\colO}{\mathscr{O}}
\safemath{\colP}{\mathscr{P}}
\safemath{\colQ}{\mathscr{Q}}
\safemath{\colR}{\mathscr{R}}
\safemath{\colS}{\mathscr{S}}
\safemath{\colT}{\mathscr{T}}
\safemath{\colU}{\mathscr{U}}
\safemath{\colV}{\mathscr{V}}
\safemath{\colW}{\mathscr{W}}
\safemath{\colX}{\mathscr{X}}
\safemath{\colY}{\mathscr{Y}}
\safemath{\colZ}{\mathscr{Z}}

\safemath{\opA}{\mathbb{A}}
\safemath{\opB}{\mathbb{B}}
\safemath{\opC}{\mathbb{C}}
\safemath{\opD}{\mathbb{D}}
\safemath{\opE}{\mathbb{E}}
\safemath{\opF}{\mathbb{F}}
\safemath{\opG}{\mathbb{G}}
\safemath{\opH}{\mathbb{H}}
\safemath{\opI}{\mathbb{I}}
\safemath{\opJ}{\mathbb{J}}
\safemath{\opK}{\mathbb{K}}
\safemath{\opL}{\mathbb{L}}
\safemath{\opM}{\mathbb{M}}
\safemath{\opN}{\mathbb{N}}
\safemath{\opO}{\mathbb{O}}
\safemath{\opP}{\mathbb{P}}
\safemath{\opQ}{\mathbb{Q}}
\safemath{\opR}{\mathbb{R}}
\safemath{\opS}{\mathbb{S}}
\safemath{\opT}{\mathbb{T}}
\safemath{\opU}{\mathbb{U}}
\safemath{\opV}{\mathbb{V}}
\safemath{\opW}{\mathbb{W}}
\safemath{\opX}{\mathbb{X}}
\safemath{\opY}{\mathbb{Y}}
\safemath{\opZ}{\mathbb{Z}}
\safemath{\opZero}{\mathbb{O}}
\safemath{\identityop}{\opI}

\safemath{\veca}{\bma}
\safemath{\vecb}{\bmb}
\safemath{\vecc}{\bmc}
\safemath{\vecd}{\bmd}
\safemath{\vece}{\bme}
\safemath{\vecf}{\bmf}
\safemath{\vecg}{\bmg}
\safemath{\vech}{\bmh}
\safemath{\veci}{\bmi}
\safemath{\vecj}{\bmj}
\safemath{\veck}{\bmk}
\safemath{\vecl}{\bml}
\safemath{\vecm}{\bmm}
\safemath{\vecn}{\bmn}
\safemath{\veco}{\bmo}
\safemath{\vecp}{\bmmp}
\safemath{\vecq}{\bmq}
\safemath{\vecr}{\bmr}
\safemath{\vecs}{\bms}
\safemath{\vect}{\bmt}
\safemath{\vecu}{\bmu}
\safemath{\vecv}{\bmv}
\safemath{\vecw}{\bmw}
\safemath{\vecx}{\bmx}
\safemath{\vecy}{\bmy}
\safemath{\vecz}{\bmz}

\safemath{\veczero}{\bmzero}
\safemath{\vecone}{\bmone}
\safemath{\vecxi}{\bmxi}
\safemath{\veclambda}{\bmlambda}
\safemath{\vecmu}{\bmmu}
\safemath{\vectheta}{\bmtheta}
\safemath{\vecphi}{\bmphi}

\safemath{\matA}{\bA}
\safemath{\matB}{\bB}
\safemath{\matC}{\bC}
\safemath{\matD}{\bD}
\safemath{\matE}{\bE}
\safemath{\matF}{\bF}
\safemath{\matG}{\bG}
\safemath{\matH}{\bH}
\safemath{\matI}{\bI}
\safemath{\matJ}{\bJ}
\safemath{\matK}{\bK}
\safemath{\matL}{\bL}
\safemath{\matM}{\bM}
\safemath{\matN}{\bN}
\safemath{\matO}{\bO}
\safemath{\matP}{\bP}
\safemath{\matQ}{\bQ}
\safemath{\matR}{\bR}
\safemath{\matS}{\bS}
\safemath{\matT}{\bT}
\safemath{\matU}{\bU}
\safemath{\matV}{\bV}
\safemath{\matW}{\bW}
\safemath{\matX}{\bX}
\safemath{\matY}{\bY}
\safemath{\matZ}{\bZ}
\safemath{\matzero}{\bmzero}

\safemath{\matDelta}{\bDelta}
\safemath{\matLambda}{\bLambda}
\safemath{\matPhi}{\bPhi}
\safemath{\matSigma}{\bSigma}
\safemath{\matOmega}{\bOmega}
\safemath{\matTheta}{\bTheta}

\safemath{\matidentity}{\matI}
\safemath{\matone}{\matO}

\safemath{\rnda}{A}
\safemath{\rndb}{B}
\safemath{\rndc}{C}
\safemath{\rndd}{D}
\safemath{\rnde}{E}
\safemath{\rndf}{F}
\safemath{\rndg}{G}
\safemath{\rndh}{H}
\safemath{\rndi}{I}
\safemath{\rndj}{J}
\safemath{\rndk}{K}
\safemath{\rndl}{L}
\safemath{\rndm}{M}
\safemath{\rndn}{N}
\safemath{\rndo}{O}
\safemath{\rndp}{P}
\safemath{\rndq}{Q}
\safemath{\rndr}{R}
\safemath{\rnds}{S}
\safemath{\rndt}{T}
\safemath{\rndu}{U}
\safemath{\rndv}{V}
\safemath{\rndw}{W}
\safemath{\rndx}{X}
\safemath{\rndy}{Y}
\safemath{\rndz}{Z}

\safemath{\rveca}{\bimA}
\safemath{\rvecb}{\bimB}
\safemath{\rvecc}{\bimC}
\safemath{\rvecd}{\bimD}
\safemath{\rvece}{\bimE}
\safemath{\rvecf}{\bimF}
\safemath{\rvecg}{\bimG}
\safemath{\rvech}{\bimH}
\safemath{\rveci}{\bimI}
\safemath{\rvecj}{\bimJ}
\safemath{\rveck}{\bimK}
\safemath{\rvecl}{\bimL}
\safemath{\rvecm}{\bimM}
\safemath{\rvecn}{\bimN}
\safemath{\rveco}{\bomO}
\safemath{\rvecp}{\bimP}
\safemath{\rvecq}{\bimQ}
\safemath{\rvecr}{\bimR}
\safemath{\rvecs}{\bimS}
\safemath{\rvect}{\bimT}
\safemath{\rvecu}{\bimU}
\safemath{\rvecv}{\bimV}
\safemath{\rvecw}{\bimW}
\safemath{\rvecx}{\bimX}
\safemath{\rvecy}{\bimY}
\safemath{\rvecz}{\bimZ}

\safemath{\rvecxi}{\bmxi}
\safemath{\rveclambda}{\bmlambda}
\safemath{\rvecmu}{\bmmu}
\safemath{\rvectheta}{\bmtheta}
\safemath{\rvecphi}{\bmphi}

\safemath{\rmatA}{\bimA}
\safemath{\rmatB}{\bimB}
\safemath{\rmatC}{\bimC}
\safemath{\rmatD}{\bimD}
\safemath{\rmatE}{\bimE}
\safemath{\rmatF}{\bimF}
\safemath{\rmatG}{\bimG}
\safemath{\rmatH}{\bimH}
\safemath{\rmatI}{\bimI}
\safemath{\rmatJ}{\bimJ}
\safemath{\rmatK}{\bimK}
\safemath{\rmatL}{\bimL}
\safemath{\rmatM}{\bimM}
\safemath{\rmatN}{\bimN}
\safemath{\rmatO}{\bimO}
\safemath{\rmatP}{\bimP}
\safemath{\rmatQ}{\bimQ}
\safemath{\rmatR}{\bimR}
\safemath{\rmatS}{\bimS}
\safemath{\rmatT}{\bimT}
\safemath{\rmatU}{\bimU}
\safemath{\rmatV}{\bimV}
\safemath{\rmatW}{\bimW}
\safemath{\rmatX}{\bimX}
\safemath{\rmatY}{\bimY}
\safemath{\rmatZ}{\bimZ}

\safemath{\rmatDelta}{\bimDelta}
\safemath{\rmatLambda}{\bimLambda}
\safemath{\rmatPhi}{\bimPhi}
\safemath{\rmatSigma}{\bimSigma}
\safemath{\rmatOmega}{\bimOmega}
\safemath{\rmatTheta}{\bimTheta}

%% file: standard-macros.tex
\usepackage{amssymb}
\usepackage{amsfonts}
\usepackage{mathrsfs}
\usepackage{xspace}
\usepackage{bm}
\usepackage{fancyref}
\usepackage{textcomp}

\newenvironment{textbmatrix}{	\setlength{\arraycolsep}{2.5pt}%
								\big[\begin{matrix}}{\end{matrix}\big]%
								\raisebox{0.08ex}{\vphantom{M}}}

\def\be{\begin{equation}}
\def\ee{\end{equation}}
\def\een{\nonumber \end{equation}}
\def\mat{\begin{bmatrix}}
\def\emat{\end{bmatrix}}
\def\btm{\begin{textbmatrix}}
\def\etm{\end{textbmatrix}}

\def\ba#1\ea{\begin{align}#1\end{align}}
\def\bas#1\eas{\begin{align*}#1\end{align*}}
\def\bs#1\es{\begin{split}#1\end{split}} 
\def\bg#1\eg{\begin{gather}#1\end{gather}}
\def\bml#1\eml{\begin{multline}#1\end{multline}}
\def\bi#1\ei{\begin{itemize}#1\end{itemize}}

\newcommand{\lefto}{\mathopen{}\left}

\DeclareMathOperator{\had}{\odot}			
\DeclareMathOperator{\four}{\opF}			
\DeclareMathOperator{\Exop}{\opE}			

\DeclareMathOperator{\landauo}{\mathit{o}}

\newcommand{\Ex}[2]{\ensuremath{\Exop_{#1}\lefto[#2\right]}} 	
\newcommand{\abs}[1]{\left\lvert#1\right\rvert}		

\newcommand{\vecnorm}[1]{\lVert#1\rVert}		
\newcommand{\conj}[1]{\ensuremath{#1^{*}}} 	
\newcommand{\tp}[1]{\ensuremath{#1^{T}}} 		

\safemath{\dirac}{\delta}					
\safemath{\krond}{\dirac}					
\newcommand{\allz}[2]{\ensuremath{#1=0,1,\ldots,#2-1}}

\safemath{\upto}{\uparrow}
\safemath{\downto}{\downarrow}
\safemath{\iu}{j}							
\safemath{\ev}{\lambda}						
\safemath{\hilseqspace}{l^{2}}				
\newcommand{\banachfunspace}[1]{\setL^{#1}}	
\safemath{\hilfunspace}{\banachfunspace{2}}	
\safemath{\hilfunspacep}{\hilfunspace(\reals)}	

\safemath{\SNR}{\rho} 				
\safemath{\No}{N_0}							
\safemath{\Es}{E_s}							
\safemath{\Eb}{E_b}							
\safemath{\EbNo}{\frac{\Eb}{\No}}
\safemath{\EsNo}{\frac{\Es}{\No}}

\DeclareMathOperator{\CHop}{\ensuremath{\opH}} 
\safemath{\tvir}{\rndh_{\CHop}}				
\safemath{\tvtf}{\rndl_{\CHop}}				
\safemath{\spf}{\rnds_{\CHop}}				
\safemath{\bff}{H_{\CHop}}					

\safemath{\ircf}{r_{h}}						
\safemath{\tftvcf}{r_{s}}					
\safemath{\tfcf}{r_{l}}						
\safemath{\bfcf}{r_{H}}						

\safemath{\tcorr}{c_h}						
\safemath{\scf}{c_{s}}						
\safemath{\tfcorr}{c_{l}}					
\safemath{\fcorr}{c_{H}}						

\safemath{\mi}{I}							
\safemath{\capacity}{C}						
\safemath{\difent}{\mathsf{h}}

\newcommand{\iid}{i.i.d.\@\xspace}

\safemath{\normal}{\mathcal{N}}			
\safemath{\jpg}{\mathcal{CN}}			
\safemath{\mchain}{\leftrightarrow}		
\newcommand{\given}{\,\vert\,}				

\safemath{\dB}{\,\mathrm{dB}}
\safemath{\dBm}{\,\mathrm{dBm}}
\safemath{\Hz}{\,\mathrm{Hz}}
\safemath{\kHz}{\,\mathrm{kHz}}
\safemath{\MHz}{\,\mathrm{MHz}}
\safemath{\GHz}{\,\mathrm{GHz}}
\safemath{\s}{\,\mathrm{s}}
\safemath{\ms}{\,\mathrm{ms}}
\safemath{\mus}{\,\mathrm{\text{\textmu}s}}
\safemath{\ns}{\,\mathrm{ns}}
\safemath{\ps}{\,\mathrm{ps}}
\safemath{\meter}{\,\mathrm{m}}
\safemath{\mm}{\,\mathrm{mm}}
\safemath{\cm}{\,\mathrm{cm}}
\safemath{\m}{\,\mathrm{m}}
\safemath{\W}{\,\mathrm{W}}
\safemath{\mW}{\, \mathrm{mW}}
\safemath{\J}{\,\mathrm{J}}
\safemath{\K}{\,\mathrm{K}}
\safemath{\bit}{\,\mathrm{bit}}
\safemath{\nat}{\,\mathrm{nat}}

\safemath{\define}{\triangleq}			

\providecommand{\inner}[2]{\ensuremath{\langle#1,#2\rangle}}
\safemath{\equivalent}{\sim}
\safemath{\distas}{\sim}					
\safemath{\sdiff}{\Delta}				

\safemath{\reals}{\mathbb{R}}
\safemath{\positivereals}{\reals_{+}}
\safemath{\integers}{\mathbb{Z}}
\safemath{\posint}{\integers_{+}}
\safemath{\naturals}{\mathbb{N}}
\safemath{\posnaturals}{\naturals_{+}}
\safemath{\complexset}{\mathbb{C}}
\safemath{\rationals}{\mathbb{Q}}
\safemath{\genfield}{\mathbb{K}}
\newcommand*{\fancyrefapplabelprefix}{app}		
\newcommand*{\fancyrefthmlabelprefix}{thm}		
\newcommand*{\fancyreflemlabelprefix}{lem}		
\newcommand*{\fancyrefcorlabelprefix}{cor}		
\newcommand*{\fancyrefdeflabelprefix}{dfn}		
\newcommand*{\fancyrefproplabelprefix}{prop}		
\frefformat{vario}{\fancyrefseclabelprefix}{Section~#1}
\frefformat{vario}{\fancyrefthmlabelprefix}{Theorem~#1}
\frefformat{vario}{\fancyreflemlabelprefix}{Lemma~#1}
\frefformat{vario}{\fancyrefcorlabelprefix}{Corollary~#1}
\frefformat{vario}{\fancyrefdeflabelprefix}{Definition~#1}
\frefformat{vario}{\fancyreffiglabelprefix}{Fig.~#1}
\frefformat{vario}{\fancyrefapplabelprefix}{Appendix~#1}
\frefformat{vario}{\fancyrefeqlabelprefix}{(#1)}
\frefformat{vario}{\fancyrefproplabelprefix}{Property~#1}

%% file: defs.tex
\newcommand{\specdensity}[1]{c_{#1}}			
\newcommand{\specdensitymat}[1]{\matC_{#1}}	

\let\time\undefined
\safemath{\bandwidth}{W}		
\safemath{\duration}{D}			
\safemath{\time}{t}				
\safemath{\freq}{f}				
\safemath{\dtdf}{\dtime,\dfreq}  
\safemath{\altdtdf}{\altdtime,\altdfreq}
\safemath{\dtime}{k}				
\safemath{\dfreq}{n}				
\safemath{\altdtime}{l}				
\safemath{\altdfreq}{m}				
\safemath{\Ddtime}{\dtime}		
\safemath{\Ddfreq}{\dfreq}		
\safemath{\delay}{\tau}			
\safemath{\doppler}{\nu}			
\safemath{\maxDoppler}{\doppler_{0}}		
\safemath{\altmaxDoppler}{\widetilde{\maxDoppler}}
\safemath{\maxDelay}{\delay_{0}}			
\safemath{\altmaxDelay}{\widetilde{\maxDelay}}
\safemath{\spread}{\Delta_{\CHop}}		
\safemath{\altspread}{\widetilde{\Delta}_{\CHop}}
\safemath{\tstep}{T}				
\safemath{\fstep}{F}				
\safemath{\alttstep}{\widetilde{\tstep}}				
\safemath{\altfstep}{\widetilde{\fstep}}				
\safemath{\tfstep}{\tstep\fstep}	
\safemath{\srtfstep}{\sqrt{\tfstep}}
\safemath{\fslots}{N}			
\safemath{\tslots}{K}			
\safemath{\tfslots}{\tslots\fslots}	
\safemath{\tfsamples}{\dtime\tstep,\dfreq\fstep}	
\safemath{\WHset}{(\logon,\tstep,\fstep)}
\safemath{\WHsetsquare}{(\logon,\srtfstep,\srtfstep)}
\safemath{\spreadset}{\setD}
\safemath{\altspreadset}{\setE}
\safemath{\spreadsquareset}{\widetilde{\spreadset}}

\safemath{\dd}{\doppler,\delay}
\let\tvir\undefined
\safemath{\tvir}{h_{\CHop}}				
\safemath{\tvirp}{\tvir(\time,\delay)}	
\let\spf\undefined
\safemath{\spf}{S_{\CHop}}				
\safemath{\spfp}{\spf(\doppler,\delay)}	
\safemath{\corrtd}{R_{\CHop}}			
\safemath{\corrtdp}{\corrtd(\time,\delay)}
\safemath{\scafun}{C_{\CHop}}			
\safemath{\scafunp}{\scafun(\doppler,\delay)}	
\safemath{\ch}{h}						
\safemath{\altch}{\tilde{h}}					
\safemath{\wgn}{w}						
\safemath{\wgnp}{\wgn(\time)}			
\safemath{\wgnvec}{\vecw}				
\safemath{\chvec}{\vech}					
\safemath{\specparam}{\theta}			
\safemath{\altspecparam}{\varphi}		
\safemath{\chcorr}{r}					
\safemath{\chcorrp}{\chcorr(\time,\freq)}    
\safemath{\chspecfun}{\specdensity{}}		
\safemath{\chspecfunp}{\specdensity{}(\specparam,\altspecparam)} 
\safemath{\chspecfunmat}{\specdensitymat{}}	
\safemath{\chspecfunmatp}{\chspecfunmat(\specparam)}	
\safemath{\chspecfunentry}{c}			
\safemath{\chspecfunentryp}{\chspecfunentry_{\dfreq}(\specparam)}
\safemath{\chspecfunentrypzero}{\chspecfunentry_{0}(\specparam)}
\safemath{\approxeig}{c}				
\safemath{\approxeigp}{\approxeig[\dtdf]}

\safemath{\chset}{\setH}
\safemath{\chsetp}{\chset\lefto(\maxDelay,\maxDoppler,\uncertainty\right)}

\safemath{\logon}{g}						
\safemath{\logonp}{\logon(\time)}
\safemath{\altlogon}{\widetilde{\logon}}						
\safemath{\altaltlogon}{e}
\safemath{\altlogonp}{\altlogon(\time)}
\safemath{\altaltlogonp}{\altaltlogon(\time)}
\safemath{\slogon}{\logon_{\dtime,\dfreq}}	
\safemath{\slogonp}{\slogon(\time)}	
\safemath{\slogonset}{\left\{\logon(\time-\dtime\srtfstep)\cex{\dfreq\srtfstep\time}\right\}}
\safemath{\eftime}{T_{0}}	
 \safemath{\efband}{F_{0}} 
 \safemath{\logonf}{G} 
 \safemath{\logonfp}{\logonf(\freq)}
 \safemath{\logonalt}{f}
 \safemath{\logonaltp}{\logonalt(\time)}
 \safemath{\slogonct}{\logon_{(\alpha,\beta)}}
 \safemath{\slogonctalt}{\logon_{(\alpha',\beta')}}
 \safemath{\af}{A}						
 \safemath{\afp}{\af_{\logon}(\doppler,\delay)}	
\safemath{\setloc}{\setG}
\safemath{\setloclogon}{\widetilde{\setG}}
\safemath{\dertau}{b}
\safemath{\dernu}{a}
\safemath{\dertaup}{\dertau_{\dtdf}}
\safemath{\dernup}{\dernu_{\dtdf}}

\safemath{\const}{c}
\safemath{\constM}{\const_{M}}
\safemath{\constm}{\const_{m}}

\safemath{\dur}{\zeta}
\safemath{\rolloff}{\delta}
\safemath{\auxfun}{S}

\safemath{\inp}{x}				
\safemath{\inpp}{\inp(\time)}	

\safemath{\outp}{y}				
\safemath{\outpp}{\outp(\time)} 
 \safemath{\inpvec}{\vecx}			
 \safemath{\outpvec}{\vecy}		

 \safemath{\Pave}{P}				
 \safemath{\avP}{\Ex{}{\vecnorm{\inpvec}^{2}}\le\tslots\Pave\,\tstep}
 \safemath{\avPnorm}{(1/\tstep)\Ex{}{\vecnorm{\inpvec}^{2}}\le\tslots\Pave}
 \safemath{\setinp}{\setQ}		

 \safemath{\capacityp}{\capacity(\SNR)}	
\safemath{\minambsquare}{m_{\logon}}
\safemath{\sumambsquare}{M}
\safemath{\sumambsquarep}{\sumambsquare(\dd)}
\safemath{\maxsumambsquare}{M_\logon}
\safemath{\LBsimple}{L}
\safemath{\LBsimplep}{\LBsimple(\SNR,\logon,\tstep,\fstep,\maxDelay,\maxDoppler,\uncertainty)}
\safemath{\LBsquare}{\LBsimple_{s}}
\safemath{\LBsquarep}{\LBsquare\lefto(\SNR,\altlogon,\tfstep,\spread,\uncertainty\right)}
\safemath{\LBsquarepnew}{\LBsquare\lefto(\SNR,\logon,\tfstep,\spread,\uncertainty\right)}
\safemath{\LBund}{\LBsimple_{u}}
\safemath{\LBunds}{\widetilde{\LBsimple}_{u}}
\safemath{\LBundp}{\LBund(\SNR,\logon,\tfstep,\spread,\uncertainty)}
\safemath{\LBundsp}{\LBunds(\SNR,\logon,\tfstep,\spread,\uncertainty)}
\safemath{\ccoh}{\capacity_{coh}} 
\safemath{\cawgn}{\capacity_{\text{AWGN}}} 
\safemath{\cawgnp}{\cawgn(\SNR)} 

\safemath{\ccohp}{\ccoh(\SNR)} 
\safemath{\SNRmax}{\SNR_{max}}
\safemath{\SNRmin}{\SNR_{min}}
\safemath{\fun}{f}
\safemath{\funp}{\fun(\SNR)}
\safemath{\SNRnorm}{\gamma}
\safemath{\SNRnormMax}{\SNRnorm_{max}}
\safemath{\SNRnormMin}{\SNRnorm_{min}}

\safemath{\sumdtime}{\sum_{\dtime=-\infty}^{\infty}} 
\safemath{\sumdfreq}{\sum_{\dfreq=-\infty}^{\infty}} 
 \safemath{\limintime}{\lim_{\tslots\to\infty}}	
 \safemath{\imat}{\matidentity}
\newcommand{\spreadint}[1]{\iint #1 d\delay d\doppler}

 \newcommand{\cex}[1]{e^{\iu2\pi #1}}				
 \newcommand{\cexn}[1]{e^{-\iu2\pi #1}}			
\safemath{\euler}{\gamma}	
\safemath{\intdiscrete}{\int_{-1/2}^{1/2}}
\safemath{\intdiscretetwod}{\intdiscrete\intdiscrete}
\safemath{\hilfuntimeband}{\hilfunspace(\duration,\bandwidth)}
\newtheorem{thm}{Theorem}
\newtheorem{lem}[thm]{Lemma}

\newtheorem{dfn}[thm]{Definition}

\safemath{\uncertainty}{\epsilon}
\safemath{\uncerMax}{\uncertainty^{max}}
\safemath{\spreadMax}{\spread^{max}}

 \safemath{\interf}{p}						
 \safemath{\interfp}{\interf[\altdtdf,\dtdf]}
 \safemath{\interfmat}{\matP}						
 \safemath{\intvar}{\sigma^2_{\interf}}
 \safemath{\intvarp}{\intvar[\dtime-\altdtime,\dfreq-\altdfreq]}
 \safemath{\altwgn}{\widetilde{\wgn}}
 \safemath{\altwgnp}{\altwgn[\dtdf]}
 \safemath{\altwgnpgauss}{\altwgn_{G}[\dtdf]}
\safemath{\intvartot}{\sigma^{2}_{I}}
 \safemath{\wgnvecone}{\wgnvec_{1}}				
 \safemath{\wgnvectwo}{\wgnvec_{2}}				
 \safemath{\outpvecone}{\outpvec_{1}}                               
 \safemath{\outpvectwo}{\outpvec_{2}}                               
 \safemath{\noisesplit}{\alpha}                                              
 \safemath{\corrintinp}{\matK(\inpvec)}

\safemath{\measure}{\mu}
\safemath{\psdoned}{f}
\safemath{\psdonedp}{\psdoned(\specparam)}
\safemath{\spreadoned}{\Delta}

\safemath{\innrate}{\delta}

%% file: dmb_isit09.bbl
\begin{thebibliography}{10}
\providecommand{\url}[1]{#1}
\csname url@samestyle\endcsname
\providecommand{\newblock}{\relax}
\providecommand{\bibinfo}[2]{#2}
\providecommand{\BIBentrySTDinterwordspacing}{\spaceskip=0pt\relax}
\providecommand{\BIBentryALTinterwordstretchfactor}{4}
\providecommand{\BIBentryALTinterwordspacing}{\spaceskip=\fontdimen2\font plus
\BIBentryALTinterwordstretchfactor\fontdimen3\font minus
  \fontdimen4\font\relax}
\providecommand{\BIBforeignlanguage}[2]{{%
\expandafter\ifx\csname l@#1\endcsname\relax
\typeout{** WARNING: IEEEtran.bst: No hyphenation pattern has been}%
\typeout{** loaded for the language `#1'. Using the pattern for}%
\typeout{** the default language instead.}%
\else
\language=\csname l@#1\endcsname
\fi
#2}}
\providecommand{\BIBdecl}{\relax}
\BIBdecl

\bibitem{abou-faycal01-05a}
I.~C. Abou-Faycal, M.~D. Trott, and S.~Shamai~(Shitz), ``The capacity of
  discrete-time memoryless {Rayleigh}-fading channels,'' \emph{{IEEE} Trans.
  Inf. Theory}, vol.~47, no.~4, pp. 1290--1301, May 2001.

\bibitem{durisi08-04a}
\BIBentryALTinterwordspacing
G.~Durisi, U.~G. Schuster, H.~B{\"o}lcskei, and S.~Shamai~(Shitz),
  ``Noncoherent capacity of underspread fading channels,'' \emph{{IEEE} Trans.
  Inf. Theory}, Apr. 2008, submitted. [Online]. Available:
  \url{http://arxiv.org/abs/0804.1748}
\BIBentrySTDinterwordspacing

\bibitem{lapidoth05-07a}
A.~Lapidoth, ``On the asymptotic capacity of stationary {Gaussian} fading
  channels,'' \emph{{IEEE} Trans. Inf. Theory}, vol.~51, no.~2, pp. 437--446,
  Feb. 2005.

\bibitem{slepian76-03a}
D.~Slepian, ``On bandwidth,'' \emph{Proc. {IEEE}}, vol.~64, no.~3, pp.
  292--300, Mar. 1976.

\bibitem{etkin06-04a}
R.~H. Etkin and D.~N.~C. Tse, ``Degrees of freedom in some underspread {MIMO}
  fading channels,'' \emph{{IEEE} Trans. Inf. Theory}, vol.~52, no.~4, pp.
  1576--1608, Apr. 2006.

\bibitem{bello63-12a}
P.~A. Bello, ``Characterization of randomly time-variant linear channels,''
  \emph{{IEEE} Trans. Commun.}, vol.~11, no.~4, pp. 360--393, Dec. 1963.

\bibitem{kennedy69}
R.~S. Kennedy, \emph{Fading Dispersive Communication Channels}.\hskip 1em plus
  0.5em minus 0.4em\relax New York, NY, U.S.A.: Wiley, 1969.

\bibitem{christensen03a}
O.~Christensen, \emph{An Introduction to Frames and {Riesz} Bases}.\hskip 1em
  plus 0.5em minus 0.4em\relax Boston, MA, U.S.A.: Birkh{\"a}user, 2003.

\bibitem{wyner66-03a}
A.~D. Wyner, ``The capacity of the band-limited {Gaussian} channel,''
  \emph{Bell Syst. Tech.~J.}, vol.~45, no.~3, pp. 359--395, Mar. 1966.

\bibitem{kozek97a}
W.~Kozek, ``Matched {Weyl-Heisenberg} expansions of nonstationary
  environments,'' Ph.D. dissertation, Vienna University of Technology,
  Department of Electrical Engineering, Vienna, Austria, Mar. 1997.

\bibitem{durisi09-01b}
G.~Durisi, V.~I. Morgenshtern, and H.~B{\"o}lcskei, ``On the sensitivity of
  noncoherent capacity to the channel model,'' in preparation.

\bibitem{groechenig01a}
K.~Gr{\"o}chenig, \emph{Foundations of Time-Frequency Analysis}.\hskip 1em plus
  0.5em minus 0.4em\relax Boston, MA, U.S.A.: Birkh{\"a}user, 2001.

\bibitem{diggavi01-11a}
S.~N. Diggavi and T.~M. Cover, ``The worst additive noise under a covariance
  constraint,'' \emph{{IEEE} Trans. Inf. Theory}, vol.~47, no.~7, pp.
  3072--3081, Nov. 2001.

\bibitem{miranda00-02a}
M.~Miranda and P.~Tilli, ``Asymptotic spectra of {Hermitian} block {Toeplitz}
  matrices and preconditioning results,'' \emph{SIAM J. Matrix Anal. Appl.},
  vol.~21, no.~3, pp. 867--881, Feb. 2000.

\bibitem{matz07-05a}
G.~Matz, D.~Schafhuber, K.~Gr{\"o}chenig, M.~Hartmann, and F.~Hlawatsch,
  ``Analysis, optimization, and implementation of low-interference wireless
  multicarrier systems,'' \emph{{IEEE} Trans. Wireless Commun.}, vol.~6, no.~5,
  pp. 1921--1931, May 2007.

\end{thebibliography}
